\documentclass{article}
\usepackage{INTERSPEECH2019}
\usepackage[utf8]{inputenc}
\usepackage{booktabs}
\usepackage{tikz}
\usepackage{hyperref}
\usepackage{xcolor}
\usepackage{listings}
\usepackage{xparse}
\NewDocumentCommand{\codeword}{v}{%
\texttt{\textcolor{blue}{#1}}%
}
\makeatletter
\g@addto@macro{\UrlBreaks}{\UrlOrds}
\makeatother

\title{Pkwrap: a PyTorch Package for LF-MMI Training of Acoustic Models}
\name{Srikanth Madikeri$^{ 1}$, Sibo Tong$^{ 1,2}$, Juan Zuluaga-Gomez$^{ 1,2}$, Apoorv Vyas$^{ 1,2}$, Petr Motlicek$^{ 1}$, Hervé Bourlard$^{ 1,2}$}
\address{
  $^{1}$Idiap Research Institute, Martigny, Switzerland \\
  $^{2}$Ecole Polytechnique F\'ed\'erale de Lausanne, Switzerland }
 \email{msrikanth, stong, jzuluaga, avyas, petr.motlicek, herve.bourlard@idiap.ch}
 
\begin{document}
\maketitle

\begin{abstract}
    We present a simple wrapper that is useful to train acoustic models in PyTorch using Kaldi's LF-MMI training framework. The wrapper, called pkwrap (short form of PyTorch kaldi wrapper), enables the user
    to utilize the flexibility provided by PyTorch in designing model architectures. It exposes the LF-MMI cost function as an autograd function.
    Other capabilities of Kaldi have also been ported to PyTorch. This includes the parallel training ability
    when multi-GPU environments are unavailable and decode with graphs created in Kaldi.
    The package is available on Github at \url{https://github.com/idiap/pkwrap}.
\end{abstract}

\section{Introduction}
Kaldi is a popular free and open-source toolkit for many speech applications~\cite{povey2011kaldi}.
Among numerous functionalities from feature extraction to various modelling algorithms, it also provides the implementation for state-of-the-art acoustic modelling of speech
through Lattice-Free Maximum Mutual Information (LF-MMI) training~\cite{povey2016purely}.
This in turn has led to multiple efforts in exposing or re-implementing such functionalities in popular deep learning toolkits such as PyTorch \cite{paszke2019pytorch} with simpler interfaces.
In PyKaldi, python bindings for most of Kaldi functionality are exposed using pybind11~\cite{can2018pykaldi}.
In PyKaldi2, LF-MMI training functionality is also provided albeit with certain limitations~\cite{lu2019pykaldi2}.
One main concern is the sub-optimal memory copy of data between libraries.
In a more recent effort, PyChain provides a pure PyTorch implementation of flat-start LF-MMI training~\cite{shao2020pychain}.
As a part of Kaldi, there exist recipes implementing a port to PyTorch in \codeword{pybind11}  branch. 
 
In Kaldi, there are two methods to train the AM using the LF-MMI framework: (1) use alignments of outputs
obtained from another model (typically a Hidden Markov Model/Gaussian Mixture Model) or (2) flat-start
training with a full-biphone tree~\cite{hadian2018flat}.
We present a package that enables users to train acoustic models in PyTorch by exposing the aforementioned
functionalities of Kaldi.
Our goal of this package is twofold: (1) support both types of LF-MMI training, and (2) support parallel training by exposing Natural Gradient-Stochastic Gradient Descent (NG-SGD) in Kaldi~\cite{povey2014parallel}.
Our motivation for (1) is to provide a simple interface to the user to experiment
with different model architectures that are simpler to implement in PyTorch.
The need for (2) arises from the lack of support from PyTorch to run distributed training
in environments where multi-GPU environments are not available.
In addition, it helps test our recipes directly with the performance obtained in Kaldi.
The recipes provided with the package also facilitate decoding. 
We implemented necessary functionalities such as reading feature files, piping model outputs for decoding, reading training \codeword{egs} prepared with Kaldi, etc.

\section{Code structure}
We use \codeword{pybind11}~\cite{jakob2017pybind11} to expose Kaldi library functions, but only those necessary for acoustic model training at the moment. 
Extending it to expose all library functions is trivial. 
The codebase is split into two: (1) \codeword{src} contains C++ source files binding Kaldi functions, code necessary to convert Kalid matrices to PyTorch tensors and vice versa, (2) \codeword{pkwrap} is the python interface that users are expected to import to use the package.

\subsection{LF-MMI training}
The MMI cost function computation is exposed in PyTorch using the \codeword{KaldiChainObjfFunction} class. This is simply an implementation of the \codeword{torch.autograd.Function} interface. Through an option class that mimics the functionality of \codeword{ChainTrainingOptions} in Kaldi, we also pass parameters such as weight on cross-entropy cost
function. The actual cost is computed by Kaldi with minimal memory transfer as we only pass around pointers to memory.

Similar to Kaldi, the training recipes use an exponential function for learning rate decay.
Time-delay Neural Network (TDNN) and Factorized TDNN are implemented as \codeword{torch.nn.module}.
Currently, the context information of a model using TDNN has to manually specified.
Creation of sequences and targets for training is still handled by Kaldi even though the functionalities
are exposed through a simple subprocess call in our recipes.
However, the user can read the resulting files (typically under \codeword{egs/}).
Our goal remains to simplify this procedure further.
In the current release (as of September 2020), every recipe has a trainer script associated with it. 
This is expected to get easier with refactoring in the future either via Pytorch-lightning integration or adopting a similar design principle.

\subsection{Parallel training with NG-SGD}
In Kaldi, NG-SGD is implemented through two things: \codeword{NaturalGradientAffineTransform} and \codeword{NGState}. We follow a similar approach by implementing a PyTorch \codeword{function} that implements a linear transform given the weights, bias and inputs in the forward pass, and calls Kaldi's natural gradient function during the backward pass.

\subsection{Kaldi matrices and PyTorch tensors}
While most of our recipes only transform \codeword{Tensor} to \codeword{Matrix} or
\codeword{CudaMatrix}, we also expose functionality to do vice versa. 
This feature helps load the models trained with Kaldi in a PyTorch environment (see \codeword{load_kaldi_models} branch in the repository)~\cite{prasad2020quantization}.

\section{Experiments}
We provide recipes for four datasets at the moment:  Minilibrispeech, Switchboard, BABEL, and Librispeech (100h). While Minilibrispeech and Switchboard are monolingual setups, Babel is a multilingual setup. All three use alignments for acoustic model training. Librispeech (100h) is a flat-start training setup that doesn't require prior alignments.
The goal of our experiments is to match the performance obtained with an equivalent (in terms of model hyperparameters) recipe in Kaldi.

\subsection{Minilibrispeech}
The Minilibrispeech recipe provides a quick way to test our implementation. It is a subset of the Librispeech dataset. In our experiments, we obtained the initial alignments with HMM/GMM model trained in Kaldi. 
To train the DNN acoustic model, the training data is speed-perturbed.
Conventional MFCC features of 40 dimensions are used as input to the model.
Both Kaldi and Pkwrap use the same \codeword{egs} folder (this is typically
the folder containing files for training).
For Pytorch models, we explicitly mention the layer at which subsampling
is applied.
The subsampling factor is 3, that is, 1 in 3 frames are used for training.
For the sake of completeness here are the features in Kaldi that were also
used in Pkwrap to train the acoustic models: frame subsampling, exponential decay of learning rate, parallel training and model merging.
The model in Pkwrap was trained with Adam optimizer for two epochs.

Table~\ref{tab:minilibrespeech-results} compares the results between Kaldi and 
Pkwrap. While the difference in WER may seem large, we note that owing to the
size of the test set it is not a significant difference.

\begin{table}[t]
\caption{Comparison of results on Minilibrispeech in Word Error Rate (WER) (in \%) between Kaldi and Pkwrap for TDNN models.}
\centering
\begin{tabular}{@{}lc@{}}
\toprule
Toolkit & WER \\
\midrule
Kaldi & 20.7 \\
Pkwrap & 19.3 \\
\bottomrule
\end{tabular}
\label{tab:minilibrespeech-results}
\end{table}

\subsection{Switchboard}

Switchboard~\cite{godfrey1992switchboard} is a collection of about 2,400 two-sided telephone conversations among 543 speakers (302 male, 241 female). We evaluate over models on the Hub5'00 set (also known as eval2000 set) which is split in two smaller subsets, "Callhome" and "Switchboard" shortened as CH and SWBD, respectively. We report WER on both subsets. We use the $\sim$206 hours of speech to train acoustic models in both Kaldi~\cite{povey2011kaldi} and Pkwrap. For language modelling and decoding, we initially train a 3-gram language model with Switchboard transcripts. Additionally, we re-score the decoded \textit{eval2000 set} with a 4-gram language model built from the Fisher+Switchboard task~\cite{hannun2014deep}, which includes the transcripts from both sets (nearly 2000 hours). We follow the standard Kaldi setup (as the one proposed in~\cite{hadian2018end}), including 2-fold speed and volume data augmentation in all the reported experiments and frame sub-sampling factor of 3 to speed up the training by a factor of 2;  i-vectors are not used in any experiment (we only report the Kaldi baseline which uses i-vectors). 

The network architecture is composed of seven TDNN layers of 625 units each one without any pre-initial layer (e.g. LDA or delta-delta) for both frameworks, Kaldi and Pkwrap. We use Stochastic Gradient Descent (SGD) to train the TDNN models in Pkwrap, whereas Kaldi uses NG-SGD. The experiments ran for 4 epochs, we set a learning rate of $1e^{-4}$ and the utterances were split in chunks of 140 frames. 

\begin{table}[th]
  \caption{Comparison of word error rates (WER) (in \%) on eval2000 test set for TDNN models trained on Kaldi and Pkwrap on the 300 hours Switchboard task. The 3-gram language model is based on Switchboard, whereas the 4-gram employs Switchboard+Fisher training set transcripts.}
  \label{tab:kaldi_vs_pkwrap}
  \centering
  \begin{tabular}{ccccc}
    	& \multicolumn{4}{c}{\textbf{Hub5'00 (eval2000)}} \\
    \cline{2-5}
     	& \multicolumn{2}{c}{\textbf{3-gram}} & 
     	\multicolumn{2}{c}{\textbf{4-gram}} \\
     	\multicolumn{1}{c}{\textbf{Framework}} & \multicolumn{1}{c}{\textbf{SWBD}} & \multicolumn{1}{c}{\textbf{CH}} &
     	\multicolumn{1}{c}{\textbf{SWBD}} & \multicolumn{1}{c}{\textbf{CH}} \\
     \midrule
     Kaldi \textit{+ivec +lda} & - & - & 9.8 & 19.4 \\
     Kaldi \textit{no ivec/lda} & 13.6 & 26.4 & 12.2 & 24.6 \\
     Pkwrap \textit{no ivec/lda} & \textbf{12.8} & \textbf{25.0} & \textbf{11.6} & \textbf{23.2} \\
     \bottomrule
  \end{tabular}
\end{table}

\subsubsection{Switchboard results}

Table~\ref{tab:kaldi_vs_pkwrap} shows the results of both frameworks when training with the same number of layers, features and units i.e. the same network architecture and size. The first model corresponding to Kaldi \textit{+ivec +lda} from Table~\ref{tab:kaldi_vs_pkwrap} presents the baseline Kaldi's TDNN model for the Switchboard 300 hours task (more information in Kaldi SWBD s5c results). Differently to the baseline model, our benchmark is neither using i-vectors nor lda layers. Through experiments, we showed that Pkwrap framework outperformed relatively Kaldi's TDNN models by 5.9\% and 5.3\% in Word Error Rates (WER) for SWBD and CH subsets, respectively. With a  4-gram LM, we report a 4.9\% and 5.7\% relative improvement in WERs for SWBD and CH subsets for models trained in Pkwrap compared to  the Kaldi baseline.  

\subsection{Multilingual modelling with BABEL}
Pkwrap was tested for multilingual modelling using BABEL dataset~\cite{madikeri2020}. Datasets for several languages with limited resources were released during the BABEL project with the main goal of building keyword spotting systems.
We considered 4 BABEL languages for evaluation: Tagalog (TGL), Swahili (SWA), Zulu (ZUL), and Turkish (TUR). 
The statistics of the target languages are given in Table~\ref{tab:babelstats}. Trigram language models estimated from training data are used during testing.
\begin{table}[th]
\caption{Statistics of BABEL target languages used for  testing. Note that the Eval sets mentioned refer to the "dev" set in the official BABEL release. Only conversational speech is considered for both training and testing. All durations are calculated prior to silence removal. (PPL: perplexity)}
\centering
\begin{tabular}{@{}lcccc@{}}
\toprule
Language & Vocabulary & PPL & Train (h) & Eval (h) \\
\midrule
Tagalog & 22k & 148 & 84.5 & 10.7 \\
Swahili & 25k & 357 & 38.0 & 9.3 \\
Turkish & 41k & 396 & 77.2 & 9.8 \\
Zulu & 56k & 719 & 56.7 & 9.2\\
\bottomrule
\end{tabular}
\label{tab:babelstats}
\end{table}

In addition to the MFCCs, an online i-vector extractor of dimension 100 is trained with Kaldi and appended to MFCCs as input to the acoustic model. TDNN architecture is used with 8 hidden layers. Each hidden layer has 512 units. Frame dropping is enabled for all models. To obtain alignments to train all the TDNN models, HMM/GMM models were first trained for each language. The standard recipe from Kaldi was followed. We evaluated and compared the acoustic models trained with Pkwrap and Kaldi using the standard Kaldi decoder. The results are listed in Table~\ref{tab:babel}.

\begin{table}[h]
\caption{WERs on 4 target languages from the BABEL dataset}
\centering
\begin{tabular}{@{}lcccc@{}}
\toprule
 & Tagalog & Swahili & Turkish & Zulu \\
\midrule
Kaldi & \bf{44.3} & 37.7 & 44.7 & 52.7 \\
Pkwrap & 44.9 & \bf{36.4} & \bf{44.2} & \bf{51.4} \\
\bottomrule
\end{tabular}
\label{tab:babel}
\end{table}

\subsection{Flat-start modelling with Librispeech (100h)}
In this section we present the results of flat-start training using Pkwrap and Kaldi on the 100 hours subset of librispeech \cite{panayotov2015librispeech} data called "train-clean-100". 

As proposed in \cite{hadian2018flat}, we use full biphones to enable flat-start modeling without requiring any previously trained models. We train using $80$ dimensional filter-bank features and use a acoustic model, composed of twelve factorized-TDNN layers each with a hidden layer dimension of 1024 and bottleneck dimension of 128. Similar to Kaldi, we use NG-SGD  to train the acoustic models. For Pkwrap, we use Adam \cite{kingma2014adam} as the optimizer with a learning rate of $1e^{-3}$ that is exponentially decayed to $1e^{-5}$ over $5$ epochs of training. For Kaldi, we started with set the effective initial and final learning rates to be $3e^-4$ and $3E^-5$ respectively.

\begin{table}[th]
  \caption{Comparison of word error rates (WER) (in \%) on dev and test set for TDNNF models trained on Kaldi and Pkwrap on the 100 hours subset of Librispeech dataset. 4-gram language model was used for rescoring.}
  \label{tab:kaldi_vs_pkwrap}
  \centering
  \begin{tabular}{ccccc}

     	{Framework} & {dev-clean} & {test-clean} &
     	{dev-other} & {test-other} \\
     \midrule
     Kaldi & 5.5 & 6.0 & 19.8 & 20.8 \\
     Pkwrap & \textbf{5.1} & \textbf{5.9} & \textbf{19.1} & \textbf{20.0} \\
     \bottomrule
  \end{tabular}
\end{table}

\section{Roadmap}
Currently, the package is in its  early stages of development. In the short-term, we aim to improve the package focusing on the following things:

\begin{itemize}
    
\item \textit{Refactoring}: We aim to remove boilerplate code in the current recipes and simplify training by providing Trainer classes that can be easily extended.

\item \textit{Automatic context generation}: This is already a feature under testing. It aims to remove the hard-coded parts of the recipes to enable automatic computation of contexts for TDNN and TDNN-F based models.

\item \textit{Speaker recognition recipes}: We will also provide recipes for speaker recognition and implementation of commonly used backends such as LDA and PLDA.

\item \textit{Wrap basic functionalities}: This involves exposing commonly used scripts such as feature extraction, data folder manipulation, etc.
\end{itemize}
\section{Acknowledgement}
The research is based upon work supported by the Office of the Director of
National Intelligence (ODNI), Intelligence Advanced Research Projects
Activity (IARPA), via AFRL Contract FA8650-17-C-9116.
The views and conclusions contained herein are those of the authors and
should not be interpreted as necessarily representing the official policies or
endorsements, either expressed or implied, of the ODNI, IARPA, or the
U.S. Government. The U.S. Government is authorized to reproduce and
distribute reprints for Governmental purposes notwithstanding any
copyright annotation thereon.
The work was also partially supported by European Union’s Horizon 2020 project No. 864702 - ATCO2 (Automatic collection and processing of voice data from air-traffic communications), which is a part of Clean Sky Joint Undertaking.
\bibliographystyle{IEEEtran}
\bibliography{mybib}
\end{document}